\documentclass[aps,floats,showpacs]{revtex4}

\input epsf.tex
\epsfclipon

\newcommand{\NC}{\newcommand}
\NC{\beq}{\begin{equation}}
\NC{\eeq}{\end{equation}}
\NC{\beqa}{\begin{eqnarray}}
\NC{\eeqa}{\end{eqnarray}}
\NC{\lra}{\leftrightarrow}

\NC{\sss}{\scriptscriptstyle}
\NC{\lsim}{\mbox{\raisebox{-.6ex}{~$\stackrel{<}{\sim}$~}}}
\NC{\gsim}{\mbox{\raisebox{-.6ex}{~$\stackrel{>}{\sim}$~}}}

\def\adot{\frac{\dot a}{a}}

\def\ndot{\frac{\dot n}{n}}
\def\fdot{\frac{\dot f}{f}}
\def\addot{\frac{\ddot a}{a}}

\def\fddot{\frac{\ddot f}{f}}
\def\ap{\frac{a'}{a}}

\def\np{\frac{n'}{n}}

\def\barko{\bar k_1^{-1}}
\def\barkos{\bar k_1^{-2}}

\begin{document}

\title{Can codimension-two branes solve the cosmological constant problem?}
\author{J\'er\'emie Vinet$^{\dagger}$ and James M. Cline$^{\dagger}$ }
\affiliation{\\
$^\dagger$ Physics Department, McGill University, Montr\'eal, Qu\'ebec, Canada H3A 2T8}
\email{vinetj@physics.mcgill.ca,jcline@physics.mcgill.ca}

%%%%%%%%%%%%%%%%%%%%%%%%%%%%%%%%%%%%%%%%%%%%%%%%%%%%%%%%%%%%%%%%%%%%%%%%%%%%
\date{\today}

\begin{abstract}{It has been suggested that codimension-two braneworlds might 
naturally explain the vanishing of the 4D effective cosmological constant, due
to the automatic relation between the deficit angle and the brane tension. To
investigate whether this cancellation happens dynamically, and within the
context of a realistic cosmology, we study a codimension-two braneworld with
spherical extra dimensions compactified by magnetic flux.   Assuming Einstein
gravity, we show that when the brane contains matter with an arbitrary equation
of state, the 4D metric components are not regular at the brane, unless the
brane has nonzero thickness.  We construct explicit 6D solutions with thick
branes, treating the brane matter as a perturbation, and find that the universe
expands consistently with standard Friedmann-Robertson-Walker (FRW) cosmology.  The relation between the
brane tension and the bulk deficit angle becomes $\Delta=2\pi G_6(\rho-3 p)$
for a general equation of state.  However, this relation does not imply a
self-tuning of the effective 4D cosmological constant to zero; perturbations of
the brane tension in a static solution lead to deSitter or anti-deSitter
braneworlds. Our results thus confirm other recent work showing that
codimension-two braneworlds in nonsupersymmetric Einstein gravity do not lead
to a dynamical relaxation of the cosmological constant, but they leave open the possibility
that supersymmetric versions can be compatible with self-tuning.}
\end{abstract}
%%%%%%%%%%%%%%%%%%%%%%%%%%%%%%%%%%%%%%%%%%%%%%%%%%%%%%%%%%%%%%%%%%%%%%%%%%%%


\maketitle

\section{Introduction}

The impressive correspondence between the ever-increasing amount of  
high-precision cosmological data \cite{SN1a,WMAP} and the theoretical predictions of the 
$\Lambda$-Cold Dark Matter ($\Lambda$CDM) model is undeniably one of the greatest  success 
stories of modern physics.  Casting a shadow over this success
however is the nagging  fact that there is to date no satisfactory
understanding of just what the dark energy  ($\Lambda$) component is, and why
its properties are what they are.   Working under the assumption that
this term is a cosmological  constant due to vacuum energy, naive estimates
of its size come out a  staggering $60-120$ orders of magnitude too large
(the number depending on one's beliefs about supersymmetry), which until
recently led most physicists to suspect that there must exist some as-yet
undiscovered mechanism which exactly cancels its effect.  The fact that
observations now tell us that this term is {\it not} exactly zero, but
nevertheless extraordinarily small by particle physics standards, means that 
finding a theoretical framework where we can understand its scale and origin
has become a pressing challenge for the theoretical physics community.  

One promising line of attack on the cosmological constant problem, which has recently
garnered renewed interest, centers around ideas which arose some twenty  years
ago with the work of Rubakov and Shaposhnikov \cite{Rubakov:1983bz}.   These
authors realized that in certain six-dimensional models, the expansion rate
of  the three large spatial dimensions is an integration constant  and thus
independent of the vacuum energy.  Several papers followed, e.g. 
\cite{Randjbar-Daemi:1985wg}-\cite{Gibbons:1986wg}, 
which studied features such as stability and cosmology in these models.  

Lately, the advent of braneworld scenarios \cite{ADD}-\cite{RSII} has 
rekindled interest in these ideas as possible ways of addressing the 
cosmological constant problem \cite{Chen:2000at}-\cite{deCarlos:2003nq}.  In
the new perspective, the old  six dimensional models can be recast as
codimension-two braneworlds that have the interesting property that the brane
tension, rather than leading to expansion, induces a deficit angle in the
bulk.  Thus the Hubble expansion measured by observers on the brane would be
insensitive to the vacuum energy associated with any field theory confined to
that brane.  This has led to the hope that there might exist a  mechanism
whereby if the brane tension suddenly changed, as in a phase transition,  the
deficit angle would dynamically relax to a value which cancels the tension's
contribution to the expansion rate,  thus providing at least a partial
solution  to the cosmological constant problem.  (It would be only a partial
solution since it is still necessary to do one fine-tuning of bulk 
contributions to the effective 4D cosmlogical constant.)

However there are open questions concerning this  scenario.  It is not at all
clear that such a tuning mechanism is  really possible within Einstein
gravity \cite{Nilles:2003km}-\cite{Graesser:2004xv}.   Furthermore, it
has been argued that it is 
impossible to find solutions with any matter sources on the
brane except for those with the equation of state of pure tension
\cite{Cline:2003ak,Gregory2}, which is a major obstruction to  building more
general models that would allow one to recover FRW cosmology at  late times.  

Much effort has been devoted to studying the cosmology of codimension-one
branes,  and the associated corrections to standard cosmology are well-known
\cite{BDL}-\cite{CGS}.  Lately codimension-two braneworlds have received
increasingly more attention, but most of this work has focused on  static
solutions or issues of stability 
\cite{Sundrum:1998ns}-\cite{Kehagias}, while the papers addressing
cosmology have considered  almost exclusively only pure tension branes (see,
however,  \cite{Gregory2},\cite{Navarro:2004kh}-\cite{Kaloper} for recent
proposals for generalizing the brane equation  of state, and
\cite{Schwindt:2003er} for a different perspective on the matter).  

In this paper, we argue that the apparent obstruction to finding solutions 
with a more general equation of state for codimension-two branes is not a problem
in principle; rather it stems from an unrealistic expectation that the bulk
geometry should be regular in the vicinity of the brane.  This singularity
happens to be absent for branes whose equation of state is that of pure
tension, $\rho=-p$. But in general, sources of codimension $p$ in gravity and
electrostatics are well known to give singular solutions $\sim r^{2-p}$, so it
should be no surprise to find a logarithmic singularity in the metric in the
codimension-2 case.  To study this in a controlled manner, we will regularize
the singularity by giving the brane a finite thickness $r_0$, and eventually
take the limit $r_0\to 0$.  

In terms of physics, our goal is to answer two specific questions: 
\begin{itemize}
\item Can the deficit angle dynamically relax to a value which cancels the effect 
of the brane tension?
 \item Do these models lead to FRW cosmology when the matter on the brane has an 
arbitrary equation of state?
\end{itemize}

The plan of the paper is as follows: in Sec. II, we review the argument 
behind the claim that one can only have pure tension on codimension-two delta 
function branes in Einstein gravity.  In Sec. III we introduce the thick-brane
setup we will be using to circumvent this difficulty.  In Sec. IV, we review
the known static background solution which we  will be perturbing to obtain
cosmological solutions.  We present  the linearized perturbative equations of
motion and the Friedmann equations which follow from their solutions in Section
V.   Section VI rephrases these results in terms of effective four-dimensional
quantities that are relevant to observers living on the brane, and derives the
Friedmann equations.  The reader who is interested in the main results of the
paper may wish to skip directly to this point.    We  discuss our results in
Sec. VII and give conclusions in Sec. VIII.  Full solutions to  the perturbed
equations of motion are quite lengthy, and are provided  to the interested
reader in the appendices.

It should be emphasized that supersymmetric versions of the  model we are
considering  \cite{Aghababaie:2003wz,Aghababaie:2003ar,Burgess:2004kd} may be more promising, and
represent a worthwhile subject for further research.  

%%%%%%%%%%%%%%%%%%%%%%%%%%%%%%%%%%%%%%%%%%%%%%%%%%%%%%%%%%%%%%%%%%%%%%%%%%%%%%%%%%%%%%%%%%%%%%%%%%%%
%---------------------------------------------------------------------------------------------------
%%%%%%%%%%%%%%%%%%%%%%%%%%%%%%%%%%%%%%%%%%%%%%%%%%%%%%%%%%%%%%%%%%%%%%%%%%%%%%%%%%%%%%%%%%%%%%%%%%%%

\section{The trouble with delta function branes}

It has already been pointed out in \cite{Cline:2003ak} and 
confirmed by \cite{Gregory2} that in Einstein gravity, codimension-two 
delta function branes appear to be incompatible with the matter on the 
brane having a general equation of state.  We repeat here the crux of 
the argument.
  
We start with the most general six dimensional metric with the two 
extra dimensions having axial symmetry
\beqa
ds^2 = -N^2(\bar r,\bar t)d\bar t^2+A^2(\bar r,\bar t)d\vec x^2 
+ B^2(\bar r,\bar t)d\bar r^2 + C^2(\bar r,\bar t)d\theta^2 
+ 2E(\bar r,\bar t)d\bar rd\bar t
\eeqa
where $t$ and $\vec x$ correspond to the usual four large dimensions, while 
$r$ and $\theta$ span the two dimensional internal space.  This metric can 
be put into the following form through a suitable transformation
of the coordinates $\bar r(r,t)$ and $\bar t(r,t)$:
\beqa
-n^2(r,t)dt^2+a^2(r,t)d\vec x^2 + f^2(r,t)(dr^2+r^2d\theta^2).
\eeqa
The equations relevant to our discussion are the $(tt)$ and $(xx)$ 
components of the Einstein equations, (here and in the rest of the 
text, primes stand for derivatives with respect to $r$ while dots 
represent derivatives with respect to $t$),
\beqa
&&\frac{1}{f^2}\left[\nabla^2\ln(f)+3\nabla^2\ln(a) +6
\left(\ap\right)^2\right]-\frac{1}{n^2}\left[3\left(\adot\right)^2
+\left(\fdot\right)^2+6\adot\fdot\right] = 8\pi G_6 T^t_t\nonumber\\\\
&&\left[\nabla^2\ln(f)+\nabla^2\ln(n)+2\nabla^2\ln(a)
+2\np\ap+2\left(\ap\right)^2+\left(\np\right)^2\right]\nonumber\\
&&+\frac{1}{n^2}\left[2\adot\ndot-2\addot-\left(\adot\right)^2
-4\adot\fdot-\left(\fdot\right)^2-2\fddot+2\ndot\fdot\right] = 
8\pi G_6 T^x_x
\eeqa
where $\nabla^2$ is the two-dimensional Laplacian for the 
coordinates $r$ and $\theta$.  Supposing there is a codimension-two delta function brane at the origin 
of the internal space, $T^t_t$ and $T^x_x$ contain the singular terms $-\rho(t){\delta(r)}/
({2\pi f^2(r,t) r})$ and $p(t)\delta(r)/({2\pi f^2(r,t) r})$ 
respectively.  Since in two dimensions the delta function
can be written as
\beqa
\delta(r) = 2\pi r \nabla^2\ln(r) ,
\eeqa
the singular parts of the Einstein tensor must come from the terms 
$\nabla^2\ln(f)$, $\nabla^2\ln(a)$ or $\nabla^2\ln(n)$.  It is
not immediately obvious how to interpret having 
$a(0,t)\sim r^{f_a(t)}$ or $n(0,t)\sim r^{f_n(t)}$ if we are to put the standard 
model on a 3-brane at the origin, since the metric is either vanishing or singular 
(depending on the sign of the exponent) there.
If we consider this to be a problem, the Einstein tensor only allows for 
\beqa
G^t_t|_{\rm sing} = G^x_x|_{\rm sing}
\eeqa
and we are led to conclude that codimension-two delta function branes can 
only have a stress energy tensor of the form $T^t_t = T^x_x$ (with $T^r_r=
T^{\theta}_{\theta}=0$).  Furthermore, the deficit angle is  related to an
integral over the singular part of the Einstein tensor,  and is therefore
rigidly related to the brane tension through the Einstein equations.  This
prevents us from being able to  meaningfully ask the question ``what happens
if there  is a sudden phase transition on the brane, so that the tension and
deficit  angle become detuned?'' unless we somehow regularize the
behaviour of the metric at the brane.  

One possible regularization is to give the brane a  finite thickness.  By
smoothing the singularity in the source, we remove the necessity of making the
metric components vanish or become singular at the position of the brane.  This
is the approach we will follow in this paper, since we feel that it is less
contrived than certain modifications to general relativity which could
alternatively be invoked, and it allows for a better physical understanding of
the apparent obstruction to zero-thickness branes with a general equation of
state.

Among the other possibilities that invoke modified gravity, which we
will not pursue here, is the addition  of an intrinsic curvature term to the
brane action, as in Dvali-Gabadadze-Porrati braneworlds  (see, e.g.
\cite{DGP,DGPothers}).  It has been argued that this  scenario should
arise naturally since the matter on the brane can induce a  4D kinetic term
for gravitons through loop corrections \cite{DGP}.  Such  terms can allow one
to obtain additional delta function contributions  in the equations of motion
to match with a brane where $\rho\neq -p$.   In this scenario, the deficit
angle acts on the brane like a negative  cosmological constant which, in
contrast with what happens in the absence  of the intrinsic curvature term,
is no longer forced by the equations of  motion to be equal to the brane
tension.  

A second possibility, as was shown in \cite{Gregory2}, is to add Gauss-Bonnet
terms to the bulk action, which allows one to both put  arbitrary types of
matter and recover Einstein gravity on the brane.   The result, as with the
previous option, is that the exact cancellation  between the brane tension
and deficit angle is no longer imposed by the  equations of motion.  

Recently a third proposal \cite{Navarro:2004kh,Lee:2004hh} was made, using 
the intersection of codimension-one branes rather than a simple 
``point-like'' codimension-two brane as the place where matter is 
localized. However since both references include extra terms in the 
gravitational action, it is not clear whether it is the  different geometry
or the generalized action which allows one to  generalize the brane equation
of state. 

Our view is that the above modifications are rather artificial because
there is in general no reason to expect the metric to be well-behaved in 
the vicinity of a codimension-two or higher source.  It appears to be an
accident that the singularity is merely a conical one in the case of
codimension-two source with pure tension.  We will show below that the
solution for a more general equation of state has more singular behaviour
near the brane.
%%%%%%%%%%%%%%%%%%%%%%%%%%%%%%%%%%%%%%%%%%%%%%%%%%%%%%%%%%%%%%%%%%%%%%%%%%%%%%%%%%%%%%%%%%%%%%%%%%%%
%---------------------------------------------------------------------------------------------------
%%%%%%%%%%%%%%%%%%%%%%%%%%%%%%%%%%%%%%%%%%%%%%%%%%%%%%%%%%%%%%%%%%%%%%%%%%%%%%%%%%%%%%%%%%%%%%%%%%%%

\section{Setup}

As we have explained in the previous section, to allow for a general 
equation of state on a codimension-two brane with nonsingular solutions, it is necessary in Einstein 
gravity for the brane to have finite thickness.  To study such a 
setup, we will model the branes using step function cores.  
This regularization scheme is similar to one that 
has been studied in the context of cosmic strings,  (see, e.g. \cite{Gott,
Hiscock}), with the difference that we will allow for the possibility of time 
dependent thickness.  (This time dependence cannot be ruled out a priori for thick branes 
whose matter content itself is dynamical, as will be the case here).  Furthermore, in order to find analytical solutions, we will 
treat the matter with $\rho\neq -p$ as a perturbation to a static 
pure-tension brane background. We will therefore write the metric ansatz as 
\beqa
\label{pertmetric}
%ds^2 &=& -e^{2(N_0(r)+N_1(r,t))}dt^2+a_0^2(t)e^{2(A_0(r)+A_1(r,t))}d\vec x^2 
%+ (1+B_1(r,t))^2 dr^2\nonumber\\
%&&+e^{2(C_0(r)+C_1(r))}d\theta^2+2E_1(r,t)\,dr\,dt.
ds^2 = -n(r,t)^2dt^2+a(r,t)^2d\vec x^2+b(r,t)^2dr^2+c(r,t)^2d\theta^2+2E(r,t)drdt
\eeqa
with
\beqa
n(r,t) &=& e^{N_0(r)+N_1(r,t)};\qquad a(r,t) = a_0(t)e^{A_0(r)+A_1(r,t)};\qquad b(r,t) = 1+B_1(r,t);\nonumber\\
c(r,t) &=& e^{C_0(r)+C_1(r,t)};\qquad E(r,t) = E_1(r,t).
\eeqa
The bulk stress-energy content consists of a 6D cosmological constant $\Lambda_6$ and 
a two-form field $F_{ab}$, so that the bulk action can be written as\footnote{Note that 
we will be using MTW \cite{MTW} conventions throughout the paper.}
\beqa
S_{bulk} = \int d^6x\sqrt{-g}\left(\frac{{\cal R}}{16\pi G_6}-\Lambda_6-\frac{1}{4}F^{ab}F_{ab}\right).
\eeqa
Although there exist solutions for both positive and negative values of $\Lambda_6$ \cite{Cline:2003ak}, we will 
take it to be positive only for the rest of the paper.  We assume that the only nonvanishing component of the vector 
potential is $A_{\theta}(r,t)=A_{\theta}^{(0)}(r)+A_{\theta}^{(1)}(r,t)$.  
For the sake of generality, we include a possible perturbation of the 6D cosmological constant, 
$\Lambda_6 \rightarrow \Lambda_6+\delta\Lambda_6$.  
The background flux from the gauge field is necessary for getting a simple
spherical compactification of the two extra dimensions.
The full stress-energy tensor is taken to be of the form
\beqa
T^a_b(r,t)=t^a_b(r,t)+\theta(r_0(t)-r)S^a_b(r,t)+\theta(r-r_*(t)){S_*}^a_b(r,t).
\eeqa
Here, $r_0(t)$ and $r_*(t)$ are the radial positions of the edges of the thick 3-branes centered around the poles of 
the compact internal space, $t^a_b$ refers to the bulk content, and $S^a_b$ is the core stress 
energy, 
given by
\beqa
S^t_t &=& -\sigma-\rho;\qquad S^x_x = -\sigma+p;\qquad
S^r_r = 0+p^r_r;\nonumber\\
S^{\theta}_{\theta} &=& 0 + p^{\theta}_{\theta};\qquad
S^r_t = 0+p^r_t;\qquad
S^t_r = 0+p^t_r;\nonumber\\
{S_*}^t_t &=& -\sigma-\rho_*;\qquad{S_*}^x_x = -\sigma+p_*\qquad
{S_*}^r_r = 0+{p_*}^r_r;\nonumber\\
{S_*}^{\theta}_{\theta} &=& 0 + {p_*}^{\theta}_{\theta};\qquad
{S_*}^r_t = 0+{p_*}^r_t;\qquad
{S_*}^t_r = 0+{p_*}^t_r.
\label{coreS}
\eeqa 
We treat the time dependence of the thickness as a perturbation, 
so that $r_0(t) = r_0+\Delta r_0(t)$, $r_*(t) = r_*+\Delta r_*(t)$ and 
\beqa
\theta(r_0(t)-r) &=& \theta(r_0-r)+\delta(r_0-r)\Delta r_0(t) +{\cal O}(\Delta r_0^2) \\
\theta(r-r_*(t)) &=& \theta(r-r_*)-\delta(r-r_*)\Delta r_*(t) +{\cal O}(\Delta r_*^2) .
\eeqa
so that effectively, we can write the stress-energy tensor as
\beqa
t^a_b+s^a_b+{s_*}^a_b
\eeqa
with, e.g.,
\beqa
s^t_t &=& -\sigma\theta(r_0-r) + \left[-\rho\theta(r_0-r)-\sigma\delta(r_0-r)\Delta r_0(t)\right]\\
s^i_i &=& -\sigma\theta(r_0-r) + \left[p\theta(r_0-r)-\sigma\delta(r_0-r)\Delta r_0(t)\right]
\eeqa
and similarly for all other terms.  

Here $\sigma$ represents the tension of the regularized brane, and
$\rho,p$ represent contributions from ordinary matter on the 
standard-model (SM) brane, while starred quantities refer to matter on a hidden
brane which is antipodal to the SM brane on the two-sphere bulk.
All variables except $\sigma$ are functions of both $r$ and $t$.  The 
appearance of off-diagonal terms in the perturbative stress-energy tensor 
might be surprising at first, but as we will explain, they must be 
included, since they cannot be made to vanish in a coordinate invariant 
manner.  

The subscripts on the metric and gauge field perturbations indicate their 
order in a perturbative series in powers of $\rho$.\footnote{As will be
seen in a subsequent section, consistency of the perturbative ansatz with 
coordinate invariance requires that some 
perturbations be ${\cal O}(\rho^{3/2})$ rather than 
${\cal O}(\rho)$.}  We will furthermore assume that time derivatives 
of the perturbations are of ${\cal O}(\rho^{3/2})$, which is implied by
the usual law for conservation of energy $\dot \rho \sim (\dot a/a) \rho 
\sim \rho^{3/2}$.  The framework is thus the same one that was used in 
the context of Randall-Sundrum cosmology in \cite{CF,CV}.  
%%%%%%%%%%%%%%%%%%%%%%%%%%%%%%%%%%%%%%%%%%%%%%%%%%%%%%%%%%%%%%%%%%%%%%%%%%%%%%%%%%%%%%%%%%%%%%%%%%%%
%---------------------------------------------------------------------------------------------------
%%%%%%%%%%%%%%%%%%%%%%%%%%%%%%%%%%%%%%%%%%%%%%%%%%%%%%%%%%%%%%%%%%%%%%%%%%%%%%%%%%%%%%%%%%%%%%%%%%%%

\section{The background}

In this section we will generalize the  static ``football shaped'' solution  \cite{Carroll,Navarro1}
to the case of branes with nonvanishing thickness.  (This will serve as the background around which
we will later perturb by adding cosmological matter on the branes.)  We will show that in the limit
where the brane thickness vanishes, we recover the  solution for delta function branes.  This might
at first seem like an obvious outcome, but in fact it does not necessarily happen, in the context of
codimension-two objects in Einstein gravity  \cite{Geroch}.  In the present model, the extra
dimensions are compactified  on a two-sphere, with  pure-tension branes positioned at the poles. 
The only  effect of the branes on the bulk is to induce a deficit angle so that the  internal space
effectively looks like a sphere with a wedge taken out, hence the name ``football-shaped'' (the
European reader may prefer to think in terms of a rugby ball).  We will refer to the interior
region of the thick brane as the ``core,'' and the exterior as the bulk.  

%%%%%%%%%%%%%%%%%%%%%%%%%%%%%%%%%%%%%%%%%%%%%%%%%%%%%%%%%%%%%%%%%%%%%%%%%%%%%%%%%%%%%%%%%%%%%%%%%%%%
%---------------------------------------------------------------------------------------------------
%%%%%%%%%%%%%%%%%%%%%%%%%%%%%%%%%%%%%%%%%%%%%%%%%%%%%%%%%%%%%%%%%%%%%%%%%%%%%%%%%%%%%%%%%%%%%%%%%%%%

\subsection{Core solutions}

We first consider static solutions of the unperturbed Einstein 
and gauge field equations
\beqa
8\pi G_6(\sigma+\Lambda_6)+4\pi G_6e^{-2C_0}{A_{\theta}^{(0)}}'^2+3A_0''+C_0''
+6A_0'^2+C_0'^2+3A_0'C_0' = 0\\
\label{rreq}
N_0''-A_0''+N_0'^2-3A_0'^2+C_0'(N_0'-A_0')+2A_0'N_0' = 0\\
8\pi G_6\Lambda_6 -4\pi G_6e^{-2C_0}{A_{\theta}^{(0)}}'^2+3A_0'(A_0'+N_0'+C_0')+C_0'N_0' = 0\\
\label{ththeq}
3A_0''+N_0''+3A_0'^2+N_0'^2-C_0'N_0'-3A_0'C_0' = 0\\
\label{econ}
\sigma(3A_0'+N_0') = 0\\
{A_{\theta}^{(0)}}''+{A_{\theta}^{(0)}}'(3A_0'+N_0'-C_0') = 0
\eeqa
 in the interior (core) of the
thick brane centered at $r=0$. 
In the absence of any matter perturbation,  conservation of energy 
implies eq.\ (\ref{econ}).
Once this condition is imposed, combining the $(rr)$ and 
$(\theta\theta)$ Einstein equations (\ref{rreq}) and (\ref{ththeq}) 
gives $A_0'=N_0'=0$, and the system 
of equations is solved by
\beqa
A_0(r) &=& 0;\qquad N_0(r) = 0;\qquad e^{C_0(r)} = \frac{1}{k}\sin(kr);
\nonumber\\A_{\theta}^{(0)}(r) &=& 
\frac{\beta }{k^2}\left(\cos(kr)-1\right) 
\eeqa
with 
\beq 
\label{corerels}
	k^2 = 8\pi G_6(\sigma+\beta ^2);\qquad \beta ^2 = 2\Lambda_6.
\eeq
   In
addition  to this static solution, there exist solutions where the brane
worldvolume is deSitter or anti-deSitter space if we relax the tuning between
the gauge field and cosmological  constant \cite{Cline:2003ak}.  In  the
context of a related supersymmetric model, it has been shown 
\cite{Aghababaie:2003wz,Aghababaie:2003ar} that the flat brane solution is
actually singled  out by the dilaton equation of motion.  Nothing forces us
to choose the flat brane solution, but we do so for  simplicity.  It should also 
be noted that our background assumes both branes have the same tension.  While there do 
exist solutions with different tensions on the two branes, these include warping and cannot 
be put in closed form, so the choice to expand around the simple unwarped 
background is also made for simplicity.  Since our framework will allow us to perturb these two 
unnecessary but simplifying tunings (flat branes and equal tensions), our final results will 
not be affected by this choice.  

We assume that the system is symmetric around the equator of the extra
dimensions.  Thus there is a brane also at the south pole, whose coordinate
position is taken to be $r=\pi/k - \phi$. The solution in the core for the
second brane is 
\beqa
A_0(r) &=& 0;\qquad
N_0(r) = 0;\qquad
e^{C_0(r)} = \frac{1}{k}\sin(k(r+\phi));\nonumber\\
A_{\theta}^{(0)}(r) &=& \frac{\beta }{k^2}
\left(\cos(k(r+\phi))+1\right)
\eeqa
where the difference in sign relative to the solution in the upper hemisphere is 
necessary for the gauge field to vanish at both poles.  The phase $\phi$
is an integration constant to be determined by matching the core
solutions smoothly to the bulk solutions, which we now consider.

%%%%%%%%%%%%%%%%%%%%%%%%%%%%%%%%%%%%%%%%%%%%%%%%%%%%%%%%%%%%%%%%%%%%%%%%%%%%%%%%%%%%%%%%%%%%%%%%%%%%
%---------------------------------------------------------------------------------------------------
%%%%%%%%%%%%%%%%%%%%%%%%%%%%%%%%%%%%%%%%%%%%%%%%%%%%%%%%%%%%%%%%%%%%%%%%%%%%%%%%%%%%%%%%%%%%%%%%%%%%

\subsection{Bulk solutions}

In the bulk, the static solutions look like those for
core,\footnote{In the bulk, we cannot exclude the possibility 
of warping, {\it i.e.,} nontrivial $N_0(r)$ and $A_0(r)$.  However, we will content 
ourselves with perturbing around the unwarped solutions which, in contrast to 
the warped ones, can be put into closed form.} except for the difference that 
$\sigma=0$.  The bulk solutions are therefore
\beqa
e^{\bar C_0(r)}&=& \frac{1}{\bar k_1} \sin(\bar k(r+\bar\phi))\\
\bar A_{\theta}^{(0)}(r) &=& \bar\beta\frac{1}{\bar k_1\bar k}\cos(\bar k(r+\bar\phi))-\bar \beta_1 
\eeqa
Demanding that the metric and gauge field as well as their first derivatives 
be continuous across the junctions located at $r=r_0$ and $r=\pi/k-r_0-\phi\equiv r_*$ 
(assuming that both branes have the same thickness), 
and using the fact that 
\beq
\label{bulkrels}
\bar \beta ^2 = \beta ^2=2\Lambda_6;\qquad  \bar k^2 = 8\pi G_6 \beta ^2
\eeq
(c.f.\ eq.\ (\ref{corerels})), we find the following matching conditions from
the $r=r_0$ junction:
\beqa
\label{k1sq}
\barkos &=& \frac{\sin(kr_0)^2}{k^2}+\frac{\cos(kr_0)^2}{\bar k^2}\\
\label{phieqnU}
\tan(\bar k (r_0 + \bar\phi)) &=& \frac{\bar k}{k}\tan(kr_0)\\
\bar \beta_1 &=& \frac{\beta }{k^2}\left[1-\cos(kr_0)\left(1-\frac{k^2}
{\bar k^2}\right)\right].
\eeqa
In the southern hemisphere, the bulk solution has the same form as for
the northern one, except that a different constant of integration
$\bar\beta_1^*$ must be used in place of $\bar\beta_1$ in the solution for the
gauge field.  At the $r=r_*$ junction we obtain
\beqa
\barkos &=& \frac{\sin(kr_0)^2}{k^2}+\frac{\cos(kr_0)^2}{\bar k^2}\\
\label{phieqnD}
\tan\left(\frac{\bar k}{k}\pi-\bar kr_0-\bar k\phi+\bar k\bar\phi\right) 
&=& -\frac{\bar k}{k}\tan(kr_0)\\
\bar \beta_1^* &=& -\frac{\beta }{k^2}\left[1-\cos(kr_0)\left(1-\frac{k^2}
{\bar k^2}\right)\right].
\eeqa
The matching of the bulk gauge field solutions to both cores
requires that $\bar\beta_1^* = -\bar\beta_1$, which makes the gauge field
discontinuous at the equator.  If the gauge field couples to matter, 
such a discontinuity is physically sensible only if
the solutions in the two hemispheres are
related to each other by a single-valued gauge transformation at the equator,
which leads to the quantization 
condition (see \cite{Aghababaie:2003wz,Navarro2})  
\beqa
\label{quant}
\bar \beta_1 = 
\frac{n}{2g}
\eeqa
where $n$ is an integer and $g$ is the $U(1)$ coupling constant. 

Finally, we can use 
the relations (\ref{phieqnU}) and (\ref{phieqnD}) to show that 
\beqa
\phi = 2\bar\phi +\frac{\pi}{k}+\frac{m\pi}{\bar k}
\eeqa
where $m$ is in principle an arbitrary integer, but the choice $m=-1$ ensures 
that both $\phi$ and $\bar\phi$ vanish when $\sigma$ vanishes.

Let us now show that the usual relation between 
the deficit angle and 
the brane tension emerges 
in the $r_0 \rightarrow 0$ limit of the above solution.  
Since $kr_0$ remains finite, the brane 
tension $\sigma$  scales like $1/r_0^2$.  The effective four-dimensional 
tension is found by integrating over the volume of the extra 
dimensions (see Sec. VI):
\beqa
\sigma^{(4)} = \frac{2\pi\sigma}{k^2}\left(1-\cos(kr_0)\right)
\eeqa 
so that as $r_0\rightarrow 0$, 
\beqa
4G_6\sigma^{(4)} = (1-\cos(kr_0)).
\eeqa
We must now consider how the deficit angle should be defined in 
the case we are considering, where the brane is thick.  From the bulk point of 
view, the radial distance from the brane at $r=R$ is $R-r_0$.  The circumference 
of a circle of radius $R$ is  $2\pi c(R,t)$, while the circumference 
of the brane is $2\pi c(r_0,t)$.  If there is no matter on the brane, so that 
the internal space is perfectly spherical, we would expect that as 
$r_0\rightarrow 0$ and $R\rightarrow 0$, 
\beqa
2\pi[c(R,t)-c(r_0,t)] = 2\pi(R-r_0).
\eeqa
On the other hand, if there is matter on the brane, it will modify 
the previous relation to read
\beqa
2\pi(c(R,t)-c(r_0,t)) = 2\pi(R-r_0)\left(1-\frac{\Delta}{2\pi}\right).
\eeqa
Thus we can define the deficit angle as 
\beqa
\Delta \equiv 2\pi\lim_{R\rightarrow 0}\left[\lim_{r_0\rightarrow 0} 1-\frac{c(R,t)-c(r_0,t)}{R-r_0}\right].
\eeqa
To lowest order, this simply leads to 
\beqa
\Delta &=& 2\pi\left(1-\lim_{R\rightarrow 0} \left[\lim_{r_0\rightarrow 0}
\frac{\barko\sin(\bar k(R+\bar\phi))}{R}\right]
\right)\\
&=& 2\pi\left(1-\lim_{R\rightarrow 0} 
\frac{{\cos(kr_0)}\sin(\bar kR)}{{\bar k}\,R}
\right)\\
&=& 2\pi\left(1-\cos(kr_0)\right)
\eeqa
The expected relation between the deficit angle and brane tension follows:
\beqa
\label{deficit}
\Delta = 8\pi G_6\sigma^{(4)}
\eeqa
We have therefore shown that this thick brane model consistently yields  known
results for a delta function brane in the limit of vanishing  thickness.  In
other models of localized stress energy, it could happen that the ($rr$) and
($\theta\theta$) components of the stress-energy tensor do not vanish in this
limit, in contrast to the stress tensor for a brane. In such a case, the
relation (\ref{deficit}) would no longer be guaranteed \cite{Geroch}.  Below we
will demonstrate another means by which the relation (\ref{deficit}) can be
relaxed: adding brane matter with a different equation of state generalizes the
form of (\ref{deficit}).  This is derived for thick branes, and it will be
shown that the thin-brane limit is singular for these solutions.
This is probably why departures from (\ref{deficit}) have not been pointed out 
before, to our knowledge.

%%%%%%%%%%%%%%%%%%%%%%%%%%%%%%%%%%%%%%%%%%%%%%%%%%%%%%%%%%%%%%%%%%%%%%%%%%%%%%%%%%%%%%%%%%%%%%%%%%%%
%---------------------------------------------------------------------------------------------------
%%%%%%%%%%%%%%%%%%%%%%%%%%%%%%%%%%%%%%%%%%%%%%%%%%%%%%%%%%%%%%%%%%%%%%%%%%%%%%%%%%%%%%%%%%%%%%%%%%%%

\section{Perturbations}

It would be very interesting if an inflating solution, which started out with
some relation between the brane tension and deficit angle differing from  that
required for a static solution, would dynamically relax to the static
solution.  The demonstration of such behaviour would be a convincing step
toward solving the cosmological constant problem.  It would also suggest that
cosmology should be modified for ordinary matter on the brane---given that
there is a drastic modification when the source on the brane corresponds to
vacuum energy. To explore these issues, we want to add possibly time-dependent
perturbations $\rho$ and $p$  to the  stress-energy of the brane and see how
the geometry responds.  A physical example where this would be relevant is a
phase transition on the brane, where part of the tension gets converted
instantaneously into radiation.

Before considering the perturbed equations of motion, it is useful 
to find a set of coordinate invariant quantities associated with the
metric perturbations in (\ref{pertmetric}).
Considering a small reparametrization of  $r$ and $t$ such that 
\beqa
r&\rightarrow& r + f(r,t);\qquad
t\rightarrow t + h(r,t),
\eeqa
where $f$ and $h$ are assumed to be of ${\cal O}(\rho)$, the perturbations 
transform as\footnote{We omit writing the $*$  quantities of the 
lower core, which transform in the same way.}
\beqa
N_1 &\rightarrow& N_1;\qquad
A_1 \rightarrow A_1;\qquad
B_1 \rightarrow B_1+f';\qquad\qquad\!\!
C_1 \rightarrow C_1+C_0'f;\nonumber\\
A_{\theta}^{(1)}\!\!\! &\rightarrow& A_{\theta}^{(1)}+{A_{\theta}^{(0)}}'f;
\qquad\qquad
E_1 \rightarrow E_1-h'+\dot f;\qquad
s^t_t \rightarrow s^t_t+\sigma\delta(r_0-r)f;\qquad
\nonumber\\
s^i_i &\rightarrow& s^i_i+\sigma\delta(r_0-r)f;\qquad
p^r_r \rightarrow p^r_r;\qquad\
p^{\theta}_{\theta} \rightarrow p^{\theta}_{\theta};\qquad
\ \ p^r_t \rightarrow p^r_t +\sigma\dot f;\nonumber\\
p^t_r &\rightarrow& p^t_r -\sigma h'.
\label{gts}
\eeqa
One can see why it is necessary to include off-diagonal terms in the 
perturbations to the core stress-energy tensor $S^a_b$, : they cannot be made to vanish in every frame.  
Moreover to have a consistent perturbative series in powers 
of $\rho$,  $E_1, p^r_t, p^t_r$ 
and $h$ are ${\cal O}(\rho^{3/2})$, while all other terms are of order 
$\rho$.

{}From the form of the gauge transformations, we can deduce the following 11 
gauge-invariant variables:
\footnote{Counting degrees of freedom, the 11 can be understood as 
5 from the 
metric, 6 from the stress energy tensor, plus $F_{t\theta}$ and $F_{r\theta}$, 
minus 2 from redefining $r$ and $t$.}
\beqa
Z &=& N_1' - A_1';\qquad\quad
W = 3A_1'+N_1';\qquad
X = \frac{C_1'}{C_0'}-B_1-\frac{C_0''}{C_0'^2}C_1;\nonumber\\
&& Y = {A_{\theta}^{(1)}}'-{A_{\theta}^{(0)}}'(B_1+C_1);\qquad\qquad
U = \dot A_{\theta}^{(1)}-\frac{{A_{\theta}^{(0)}}'}{C_0'}\dot C_1;\nonumber\\
\tilde\rho &=& \rho + \delta(r_0-r)\left[\frac{\sigma }{C_0'}C_1+\sigma\Delta r_0(t)\right] ;\qquad
\tilde p = p -\delta(r_0-r)\left[\frac{\sigma }{C_0'}C_1+\sigma\Delta r_0(t)\right];\nonumber\\
\tilde p^r_t &=&p^r_t-\sigma \frac{\dot C_1}{C_0'};\qquad
\tilde p^t_r = p^t_r+p^r_t-\sigma E_1;\quad
\tilde p_5 = p^r_r;\qquad
\quad \tilde p_6 = p^{\theta}_{\theta}-p^r_r .
\eeqa
In terms of these variables, the equations of motion at first
nontrivial order in the perturbation are\footnote{Here, the equations 
we have written correspond to the following combinations of the Einstein 
and gauge field equations of motion: \ref{TTminXX} is $(tt)-(xx)$, \
\ref{THminRR} is $(rr)-(\theta\theta)$, \ref{FEOM6} is  
$F^{\theta b}_{;b}$, \ref{RR} is the (rr) Einstein equation, \ref{OTHER} is 
$-1/4 (tt) -3/4 (xx) -3/4 (rr) +3/4 (\theta\theta)$, \ref{CONS5} is a combination of 
$S^a_{r;a}$ and $F^{\theta b}_{;b}$, 
\ref{GAUGE} comes from $F_{[ab;c]}=0$, \ref{TRplusRT} is $(tr)+(rt)$, 
\ref{RT} is $(rt)$ and \ref{CONS1} is a combination of 
$S^a_{t;a}$ and $F^{\theta b}_{;b}$.}
\beqa
\label{TTminXX}
Z'+C_0'Z &=& 2\left[\frac{\ddot a_0}{a_0}
-\left(\frac{\dot a_0}{a_0}\right)^2\right]
+8\pi G(\tilde\rho +\tilde p)\\
\label{THminRR}
W'-C_0'W &=& 8\pi G \tilde p_6\\
\label{FEOM6}
Y'-C_0'Y - \beta e^{C_0}W &=& 0\\
\label{RR}
C_0'W+8\pi G\beta e^{-C_0}Y &=&3\left[\frac{\ddot a_0}{a_0}
+\left(\frac{\dot a_0}{a_0}\right)^2\right]+8\pi G(\tilde p_5-\delta\Lambda_6)\\
\label{OTHER}
C_0'X'+2X\left(C_0''+C_0'^2\right)
&+&\frac{3}{2}C_0'W-8\pi G\beta e^{-C_0}Y \nonumber\\
&=& \frac{3}{2}\left[\frac{\ddot a_0}{a_0}
+\left(\frac{\dot a_0}{a_0}\right)^2\right]
-2\pi G(\tilde\rho-3\tilde p+3\tilde p_6+4\delta\Lambda_6)\\
\label{CONS5}
\tilde p_5'-C_0'\tilde p_6 +\sigma W &=& 0\\
\label{GAUGE}
U'  -\dot Y-\beta e^{C_0}\dot X &=& 0\\
\label{TRplusRT}
\tilde p^t_r &=& 0\\
\label{RT}
3\frac{\dot a_0}{a_0}Z-C_0'\dot X 
+\frac{3}{4}(\dot Z-\dot W)&=&
8\pi G(\tilde p^r_t - \beta e^{-C_0}U)\\
\label{CONS1} 
\dot{\tilde\rho}+3\frac{\dot a_0}{a_0}(\tilde\rho+\tilde p)
&=& 
\tilde {p^r_t}'+C_0'\tilde p^r_t + \sigma\dot X
\eeqa
These are not all independent however.  It can be  shown that 
eq.\ (\ref{FEOM6}) and eq.\ (\ref{GAUGE}) can be obtained as combinations of 
the other equations in the system, and can therefore be ignored.
We have found the general solution to this reduced system of equations
in appendix A.  

The general solution to (\ref{TTminXX}-\ref{CONS1}) is complicated, but we 
are principally interested in just one aspect, the Friedmann equations
which describe the expansion of the universe for an observer on one of
the branes.  In the present construction, the Friedmann equations arise
through imposing the appropriate boundary conditions at the junctions of
the regions interior and exterior to the branes.  In terms of our
gauge-invariant variables, continuity of the functions but not of the derivatives 
is required at the interfaces, since the equations of motion
do not contain any spatial second derivatives.  Moreover we require that
all functions be nonsingular at the poles.
Finally, in what follows, we 
assume that $\tilde \rho, \tilde p, \tilde \rho_*$, $\tilde p_*$ 
are functions of time only and that 
$\tilde p_{6} = e^{2C_0(r)}{\cal P}_6(t)$ and 
$\tilde p_{*6} = e^{2C_0(r)}{\cal P}_{*6}(t)$.\footnote{The variables $\tilde p_6$ 
and $\tilde p_{*6}$ don't have a dynamical equation, so we are free to specify them 
arbitrarily in the absence of a specific model for the brane matter content.}

With this ansatz for the brane energy density, applying boundary conditions 
allows us to derive the Friedmann equations.  However, there is one last point to 
address before writing them down: 
the 6D quantities $\tilde \rho$ and $\tilde p$ we have used so far are not 
the ones an observer on the brane would identify as the energy density and 
pressure.  It is therefore necessary to define what the effective energy 
density and pressure are for a 4D observer if we want to 
compare the Friedmann equations in this model to the standard ones.

%%%%%%%%%%%%%%%%%%%%%%%%%%%%%%%%%%%%%%%%%%%%%%%%%%%%%%%%%%%%%%%%%%%%%%%%%%%%%%%%%%%%%%%%%%%%%%%%%%%%
%---------------------------------------------------------------------------------------------------
%%%%%%%%%%%%%%%%%%%%%%%%%%%%%%%%%%%%%%%%%%%%%%%%%%%%%%%%%%%%%%%%%%%%%%%%%%%%%%%%%%%%%%%%%%%%%%%%%%%%

\section{Effective four dimensional quantities}

To define the effective four dimensional quantities, we integrate 
the 6D quantities over the thickness of the brane,
\beqa
{S^{(4)}}^a_b = 2\pi\int_0^{r_0} dr\, b(r,t)\,c(r,t)\, {S^{(6)}}^a_b
\eeqa
which perturbatively leads to
\beqa
\sigma^{(4)}+\rho^{(4)} &=& 2\pi\int_0^{r_0} dr\, e^{C_0}(1+B_1+C_1)(-s^t_t)\\
-\sigma^{(4)} + p^{(4)} &=& 2\pi\int_0^{r_0} dr\, e^{C_0}(1+B_1+C_1)(s^i_i)\\
p_5^{(4)} &=& 2\pi\int_0^{r_0} dr\, e^{C_0} \tilde p_5\\
p_6^{(4)} &=& 2\pi\int_0^{r_0} dr\, e^{C_0} \tilde p_6.
\eeqa
and similarly for the other brane.\footnote{Strictly speaking, we 
should also define an effective 4D scale factor $a^{(4)} \sim 
\int dr\, b(r,t)\,c(r,t)\, a^{(6)}$.  In the present case this is not necessary
because there are no corrections to $\dot a/a$ and 
$\ddot a/a$ at leading order in the perturbation.}\ \ In order to 
compute $\rho^{(4)}$ and $p^{(4)}$, we choose a gauge and find 
$C_1$ or $B_1$ explicitly.  For this purpose it is convenient to work 
in coordinates where $B_1=E_1=0$.  
{}From the
form of the gauge transformations (\ref{gts}), it can be shown that the 4D
quantities are gauge invariant.  For example, $\delta(\sigma^{(4)}+\rho^{(4)}) = 
2\pi\int_0^{r_0} dr\, {d\over dr}( e^{C_0}  \sigma f)$ is a surface term which vanishes
because $f=0$ at $r=0$ and $\sigma = 0$ at $r=r_0^+$.

Similarly, the 4D Newton constant is related to the 6D one by dimensional reduction,
\beqa
G_6 &=& G_4\times V\\
&=& G_4 \int_0^{2\pi}d\theta\int_0^{\pi/k-\phi}e^{C_0(r)}\\
&=& G_4 \times \frac{4\pi}{k^2\bar k^2}(\bar k^2 +(k^2-\bar k^2)\cos(kr_0))
\eeqa
where we neglect corrections of ${\cal O}(\rho)$.

After gauge-fixing, the Friedmann equations and conservation of energy can 
be rewritten in terms of the effective 4D quantities (see the appendix 
for the full solutions) 
\beqa
\label{fried1}
\left(\frac{\dot a_0}{a_0}\right)^2 &=& 
\frac{8\pi G_4}{3}\left(\rho^{(4)}+\rho^{(4)}_*
+\Lambda_{\rm eff}\right)\\
\label{fried2}
\frac{\ddot a_0}{a_0}&=&\left(\frac{\dot a_0}{a_0}\right)^2  
-4\pi G_4\left(\rho^{(4)}+p^{(4)}+\rho^{(4)}_*
+p^{(4)}_*\right)\\
\label{cons1}
\dot\rho^{(4)} &=& -3\frac{\dot a_0}{a_0}\left(\rho^{(4)}+p^{(4)}\right)\\
\label{cons2}
\dot\rho^{(4)}_* &=& -3\frac{\dot a_0}{a_0}\left(\rho^{(4)}_*+p^{(4)}_*\right).
\eeqa
These constitute the main result of this paper, and show that we  indeed
recover standard cosmology on the brane, except for the constant of integration
$\Lambda_{\rm eff}$ which represents bulk  contributions to the
4D cosmological constant, and the contribution of ``dark matter'' which is on
the second brane.  Of course we also expect corrections of order $\rho^2$ and
higher which would appear at higher order in the perturbation expansion
\cite{CV}, but these are not relevant for the immediate question concerning
self-tuning.

%%%%%%%%%%%%%%%%%%%%%%%%%%%%%%%%%%%%%%%%%%%%%%%%%%%%%%%%%%%%%%%%%%%%%%%%%%%%%%%%%%%%%%%%%%%%%%%%%%%%
%---------------------------------------------------------------------------------------------------
%%%%%%%%%%%%%%%%%%%%%%%%%%%%%%%%%%%%%%%%%%%%%%%%%%%%%%%%%%%%%%%%%%%%%%%%%%%%%%%%%%%%%%%%%%%%%%%%%%%%

\section{Discussion}

To elucidate our results (\ref{fried1}-\ref{cons2}),  let us first explain the
role of the constant  $\Lambda_{\rm eff}$.  It might at first
be surprising that  our solutions contains a term that does not vanish when matter
perturbations are absent.   However, this simply reflects the fact that our
choice to  perturb around a static background solution was arbitrary, and there
exist  evolving solutions even in the absence of matter perturbations.  This
is  because we are free to choose any set of values for the parameters  of the
background solution $\sigma^{(4)}$, $\beta$ and $\Lambda_6$.\footnote{The 
quantization condition then
constrains the  values the $U(1)$ coupling can take.}  We arbitrarily tuned
$\beta^2 = 2\Lambda_6$ to ensure that the brane does not expand.   Solutions 
with  nonzero $\Lambda_{\rm eff}$ simply correspond to  using a
different set of values for the background parameters,  for which the tuning 
between the field strength and bulk cosmological constant  has been perturbed. 
The  value of the constant $\Lambda_{\rm eff}$ is thus
determined by the background around which matter perturbations are being added,
and cannot be used to subsequently tune the resulting solutions.  

We will now consider some limiting cases which provide a
consistency check on our solutions, and which  illustrate the differences
between the thick-brane model and its thin-brane limit.  

\subsection{Thin brane limit}

First we would like to consider how our results  behave as we take the thin
brane ({\it i.e.,} $r_0\rightarrow 0$) limit.  Earlier we argued that for a delta
function brane with an equation of state differing from pure tension, the
4D components of the metric have singular behaviour at the
origin.  Let us explain how this is manifested in our results.

The thin brane limit corresponds to taking $k\rightarrow\infty$, since 
the constant $kr_0$  remains finite.  This follows from  the
requirement that  the effective four dimensional quantities $\sigma^{(4)}$,
$\rho ^{(4)}$ and $p^{(4)}$
%, as well as the constant of integration ${\cal Q}_1$  
should be held fixed as we vary the brane thickness.  
We can then see from our results why the $r_0\rightarrow 0$ limit produces
singular results at the branes
unless the matter on the brane consists of pure tension.  From  the 
metric perturbations $A_1$ and $N_1$, we find that\footnote{These dominant 
terms in the thin brane limit come from the gauge invariant variable Z.} 
\beqa
\label{Ap}
A_1'(r_0,t) &=& -k\frac{G_6\left(\rho^{(4)}+p^{(4)}\right)}{\sin(kr_0)} 
+{\cal O}\left(1/k\right)\\
\label{Np}
N_1'(r_0,t) &=& 3k\frac{G_6\left(\rho^{(4)}+p^{(4)}\right)}{\sin(kr_0)} 
+{\cal O}\left(1/k\right)
\eeqa
where we have concentrated only on the brane located at the upper pole,  but
would of course find similar results for the other brane.  These diverge  in
the thin brane limit (since $k\sim 1/r_0$ as $r_0\to 0$)
unless the perturbations satisfy $p^{(4)}=-\rho^{(4)}$.
Thus it is only possible to have pure tension  on a codimension-two delta
function brane, if we (unreasonably) require that $g_{tt}$ and $g_{xx}$ be regular at the
brane.  On the other hand if we allow an arbitrary equation of state, then
$A_1'$ and $N_1'$ diverge like $1/r$ near the brane, indicating that
$A_1$ and $N_1$ go like $\ln\,r$. 

We stress again that regularity at the brane is probably too much to ask of a
solution  when a delta function source is present; rather one should
expect singularities in this case.  Therefore, we do not interpret our results
as indicating any fundamental obstruction to having general types of matter on a
codimension-two delta function brane. Rather,  eqs.\ (\ref{Ap}-\ref{Np})
imply the mildly singular behaviour
\beq
	a(r,t) \sim  r^{-G_6(\rho+p)},\quad  n(r,t) \sim r^{3G_6(\rho+p)}
\eeq
for the 4D metric components in the vicinity of the brane.

\subsection{Deficit angle}
We turn our attention now to the deficit angle which, as was previously mentioned, 
is defined by
\beqa
\Delta \equiv 2\pi\lim_{R\rightarrow 0}\left[\lim_{r_0\rightarrow 0} 1-\frac{c(R,t)-c(r_0,t)}{R-r_0}\right]
\eeqa
around the brane located at $r=0$.  Plugging our solutions $c(r,t)\approx e^{C_0(r)}(1+C_1(r,t))$ 
into this, we obtain
\beqa
\label{detuneddeficit}
\Delta &=& 2\pi G_6(4\sigma^{(4)}+\rho^{(4)}-3p^{(4)}) 
\eeqa
The same procedure performed around the other brane leads to
\beqa
\label{detuneddeficitstar}
\Delta_* &=& 2\pi G_6(4\sigma^{(4)}+\rho_*^{(4)} -3p_*^{(4)})
\eeqa
We see that the deficit angle indeed responds to changes in the stress-energy
on the brane; moreover in the familiar case where 
$p^{(4)} =-\rho^{(4)}$ (pure tension), the relation reduces to its expected form 
$\Delta = 8\pi G_6 \sigma^{(4)}_{tot}$ where $\sigma^{(4)}_{tot}= \sigma^{(4)}+\rho^{(4)}$.  
However, this is not enough to insure a static solution,
as can be seen from eq.\ (\ref{fried1}).  At first this looks quite mysterious:
in the pure tension case, why is the effect of the tension not canceled by the
conical singularity, as was the case for the unperturbed solution?  The answer,
given in the next subsection, has to do with the effect of the perturbation on
the bulk fields, in particular the gauge field.

\subsection{General equations of state}

With our results we can now answer the physical question posed at the
beginning of section 5: suppose we start from a static solution which
undergoes a phase transition that converts a part of the brane's
tension into radiation---how does the geometry respond to this sudden change in
the equation of state?  The perturbation to the stress tensor can be thought
of as the sum of two pieces,
\beqa
	\rho &=& \delta\sigma + \rho_{\rm rad} \nonumber\\
	 p &=& -\delta\sigma + \frac13\rho_{\rm rad}
\eeqa
where $\delta\sigma$ is negative, and initially $\rho=0$.  According to the
Friedmann equations, the universe will start to collapse toward a big crunch,
as one would expect for an anti-deSitter solution;  
the universe behaves just as though it
had a negative cosmological constant. The conclusion actually does not require
the transition to be sudden, nor does it have to produce radiation. The same
outcome occurs even if the equation of state of the perturbation changes
arbitrarily slowly from $w=-1$ to any larger value.

For example we could consider such a transition in a more smooth way by
choosing a positive initial value of $\delta\rho$, tuning the solution  (using
the integration constant $\Lambda_{\rm eff}$) to be initially static, and imposing a
time-dependent equation of state for the perturbation, $w =  (-1+\tanh\epsilon
t)/2$.  This represents the gradual conversion of the perturbation from vacuum
energy in the past to pressureless matter in the future.  Again, the Friedmann
equations predict a collapsing universe due to the effectively negative
cosmological constant at late times.  

Similarly, if we start from an initially static solution with a negative value
of $\delta\rho$ and the opposite time dependence for the equation of state, $w
= (-1-\tanh\epsilon t)/2$,  we obtain an inflating solution at late times due
to the positive change in the brane tension.  In all cases, it is clear that
no self-tuning mechanism cancels the effect of the change in tension.

Now we address the seeming paradox: how can our results be self-consistent?
We find that the Hubble rate is sensitive to changes in the energy density
on the brane, yet the deficit angle tracks this energy density in the
way which we would hope for if there was self-tuning of the 4D cosmological 
constant to zero.  
The key point is that the  tuning between the deficit angle and brane
tension is not sufficient to lead  to a static solution; one also needs the
field strength and bulk  cosmological constant to be tuned.  In general, 
$\delta\rho$
induces a  perturbation to the field strength $F^2 \cong F_0^2+2F_0\delta F$, where  in terms of our variables, we find that 
\beqa
\delta F = \sqrt{2}e^{-C_0}Y(r,t)
\eeqa
In the static solution, the field strength had to be tuned relative to
$\Lambda_6$.  Apparently it is this tuning which is spoiled in the
presence of matter on the brane, rather than anything related to the deficit
angle.  

%%%%%%%%%%%%%%%%%%%%%%%%%%%%%%%%%%%%%%%%%%%%%%%%%%%%%%%%%%%%%%%%%%%%%%%%%%%%%%%%%%%%%%%%%%%%%%%%%%%%
%---------------------------------------------------------------------------------------------------
%%%%%%%%%%%%%%%%%%%%%%%%%%%%%%%%%%%%%%%%%%%%%%%%%%%%%%%%%%%%%%%%%%%%%%%%%%%%%%%%%%%%%%%%%%%%%%%%%%%%

\section{Summary and outlook}

In this paper we constructed a thick codimension-two braneworld model in order
to determine whether such scenarios have a self-tuning mechanism which could
help solve the  cosmological constant problem, and also to investigate the
question  of whether they allow us to recover standard cosmology.  

The good news is that the second question is answered in the affirmative. 
Codimension-two braneworlds are quite consistent with the general expectations
from decoupling and dimensional reduction: the 4D effective theory looks like
general relativity at lowest order in the density of  matter on the branes. 
Although corrections to GR will undoubtedly  appear at order $\rho^2$, we have
not tried to compute these in the present work.

On the other hand, we find an obstacle to  self-tuning of the effective 4D
cosmological constant in this model.  Sarting from a static solution, any phase
transition which changes the tension of the brane  will either lead to a
contracting or an expanding universe.  In the second case, once the matter has
redshifted away, we do not recover the original  static solution, but rather a
deSitter one, where the difference relative to the original static solution
comes from the fact that the phase transition has spoiled the original tuning
between the bulk gauge field strength and the bulk cosmological constant.  

We were motivated to study a thick brane model because of the apparent
impossibility of putting matter with a general equation of state on  a
codimension-two brane in Einstein gravity.  We find that there is actually
no prohibition; instead the metric becomes singular near the brane 
when it contains general kinds of matter.  In retrospect this is not
surprising, and the fact that pure-tension branes induce only a conical
singularity appears to be fortuitous.  Accepting the singular nature of
the metric, we derived an expression for the deficit angle with arbitrary
matter on the brane,
\beqa \Delta = 2\pi G_6\left(4\sigma^{(4)}+\rho^{(4)}-3p^{(4)}\right)
\eeqa 
which reduces to the expected one for a pure-tension brane.

Although our results show that codimension-two braneworlds in Einstein gravity
don't  solve the cosmological constant problem, the outlook may be better in 
models where there is supersymmetry in the bulk \cite{Aghababaie:2003wz,
Aghababaie:2003ar,Burgess:2004kd}.  The encouraging point in our results is
that the deficit angle {\it does} respond to arbitrary changes in the brane
stress energy in such a way as to cancel its contribution (through the 
singular part of the gravitational action).  What is needed is a symmetry to
maintain a vanishing bulk contribution to the effective 4D cosmological constant,
and it has been argued that supersymmetry can do precisely this.  In SUSY
models, the dilaton equation of motion insures that the tuning between the
gauge field and bulk cosmological constant (which appears in the guise of
a scalar field potential) required for a static solution always holds
as long as the dilaton is stabilized.   Since we have shown that it is the
detuning between the field strength and bulk cosmological constant  that leads
to non-static solutions in Einstein gravity, it seems reasonable to expect that
in  models with supersymmetry, there will indeed be a self-tuning mechanism to
cancel the effective  cosmological constant.  Checking this explicitly is the
subject of ongoing work.  

%%%%%%%%%%%%%%%%%%%%%%%%%%%%%%%%%%%%%%%%%%%%%%%%%%%%%%%%%%%%%%%%%%%%%%%%%%%%%%%%%%%%%%%%%%%%%%%%%%%%
%---------------------------------------------------------------------------------------------------
%%%%%%%%%%%%%%%%%%%%%%%%%%%%%%%%%%%%%%%%%%%%%%%%%%%%%%%%%%%%%%%%%%%%%%%%%%%%%%%%%%%%%%%%%%%%%%%%%%%%

\section{Acknowledgements}

We thank Yashar Aghababaie for useful converstations concerning flux
quantization. We also thank Cliff Burgess for patiently disabusing us of our
prejudices concerning the singular nature of the solutions, and for
emphasizing the differences between SUSY and non-SUSY models.  Finally we would 
like to thank the referee for his thouroughness, and highly constructive comments.  

JC and JV are supported in part by the National Science and Engineering
Research Council of Canada, and les Fonds Nature et Technologies of 
Qu\'ebec.  
%%%%%%%%%%%%%%%%%%%%%%%%%%%%%%%%%%%%%%%%%%%%%%%%%%%%%%%%%%%%%%%%%%%%%%%%%%%%%%%%%%%%%%%%%%%%%%%%%%%%
%---------------------------------------------------------------------------------------------------
%%%%%%%%%%%%%%%%%%%%%%%%%%%%%%%%%%%%%%%%%%%%%%%%%%%%%%%%%%%%%%%%%%%%%%%%%%%%%%%%%%%%%%%%%%%%%%%%%%%%

\appendix
\section{Perturbative solutions}
We present here general solutions to the perturbed equations of motion (\ref{TTminXX}-\ref{CONS1}).  The superscripts 
$I,II,III$ and $IV$ are meant to identify quantites associated with the different regions in the model, namely 
the upper core ($r=0$ to $r=r_0$), the upper bulk ($r=r_0$ to $r=\pi/2/\bar k -\bar\phi$), lower bulk ($r=\pi/2/\bar k-\bar\phi$ to $r=\pi/k-\phi-r_0$) and lower core ($r=\pi/k-\phi-r_0$ to $r=\pi/k-\phi$).  We have defined $\bar r\equiv r+\bar\phi$ and $\hat r \equiv r+\phi$.   We have not written the solutions for the gauge invariant variable $U(r,t)$, since 
it can be obtained trivially from the algebraic equation (\ref{RT}) once the other variables have been solved 
for.  
\beqa
Z^I &=& \frac{{\cal F}^I_1(t)}{\sin(kr)}-2\frac{\cos(kr)}{k\sin(kr)}\left[4\pi G_6(\rho+ p)+\frac{\ddot a_0}{a_0}-\left(\frac{\dot a_0}{a_0}\right)^2\right]\\
W^I &=& {\cal F}^I_2(t)\sin(kr)-\frac{8\pi G_6\sin(kr)\cos(kr)}{k^3}{\cal P}_6\\
\tilde p_5 &=& {\cal F}_3^I(t)+\frac{\sigma\cos(kr)}{k}{\cal F}_2^I(t)-\frac{(k^2+8\pi G_6\sigma)(\cos(kr)^2-\sin(kr)^2)}{4k^4}{\cal P}_6\\
Y^I &=& \frac{(k^2+8\pi G_6\sigma+2(k^2-8\pi G_6\sigma)\cos(kr)^2)\sin(kr)}{4k^5\beta}{\cal P}_6\nonumber\\
&-&\frac{(k^2-8\pi G_6\sigma)\sin(kr)\cos(kr)}{8\pi G_6\beta k^2}{\cal F}_2^I(t)\nonumber\\
&+&\frac{\sin(kr)}{8\pi G_6\beta k}\left[3\frac{\ddot a_0}{a_0}+3\left(\frac{\dot a_0}{a_0}\right)^2+8\pi G_6 {\cal F}_3^I(t)-8\pi G_6\delta\Lambda_6\right]\\
X^I &=& \frac{{\cal F}_4^I(t)}{\cos(kr)^2}+\frac{(5k^2-16\pi G_6\sigma)\cos(kr)}{6k^3}{\cal F}_2^I(t)\nonumber\\
&+&\frac{\pi G_6}{16k^6\cos(kr)^2} \left[17k^2+16\pi G_6\sigma+32\cos(kr)^2(k^2-4\pi G_6\sigma)\right.\nonumber\\
&-&\left.8\cos(kr)^4(11k^2-16\pi G_6\sigma)\right]{\cal P}_6\nonumber\\
&-&\frac{(2\cos(kr)^2-1)}{8k^2\cos(kr)^2}\left[9\frac{\ddot a_0}{a_0}+9\left(\frac{\dot a_0}{a_0}\right)^2-4\pi G_6(\rho-3 p+8\delta\Lambda_6-4{\cal F}_3^I(t))\right]\\
\tilde p^r_t &=& -3\frac{\cos(kr)}{k\sin(kr)}\frac{\dot a_0}{a_0}(\rho+ p)+\frac{d}{dt}\left[-\frac{9\sigma(1+2\cos(kr)^2)}{8k^3\sin(kr)\cos(kr)}\left(\frac{\ddot a_0}{a_0}+\left(\frac{\dot a_0}{a_0}\right)^2\right)\right.\nonumber\\
&+&\left.\frac{\sigma\pi G_6-2\cos(kr)^2(k^2-\sigma\pi G_6)}{2k^3\sin(kr)\cos(kr)}\rho-\frac{\sigma\pi G_6(1+2\cos(kr)^2)}{k^3\sin(kr)\cos(kr)}\left(\frac{3}{2} p+2{\cal F}_3^I(t)\right)\right.\nonumber\\
&+&\left.\frac{\sigma\pi G_6} {48k^7\sin(kr)\cos(kr)} \left\{-3(17k^2+16\sigma\pi G_6)+96\cos(kr)^2(k^2-4\sigma\pi G_6)\right.\right.\nonumber\\
&-&\left.\left.8\cos(kr)^4(11k^2-16\sigma\pi G_6)\right\}{\cal P}_6\right.\nonumber\\
&+&\left.\frac{(5k^2-16\pi G_6\sigma)\sigma\cos(kr)^2}{12k^4\sin(kr)}{\cal F}_2^I(t)-\frac{\sigma}{k\sin(kr)\cos(kr)}{\cal F}_4^I(t)\right]+\frac{{\cal F}_5^I(t)}{\sin(kr)}
\eeqa
\beqa
Z^{II} &=& \frac{{\cal F}^{II}_1(t)}{\sin(\bar k\bar r)}-2\frac{\cos(\bar k\bar r)}{\bar k\sin(\bar k\bar r)}\left[\frac{\ddot a_0}{a_0}-\left(\frac{\dot a_0}{a_0}\right)^2\right]\\
W^{II} &=& {\cal F}^{II}_2(t)\sin(\bar k\bar r)\\
Y^{II} &=& -\frac{ \bar k\sin(\bar k\bar r)\cos(\bar k\bar r)}{8\pi G_6\beta\bar k_1}{\cal F}_2^{II}(t)+\frac{ \sin(\bar k\bar r)}{8\pi G_6\beta\bar k_1 }\left[3\frac{\ddot a_0}{a_0}+3\left(\frac{\dot a_0}{a_0}\right)^2-8\pi G_6\delta\Lambda_6\right]\\
X^{II} &=& \frac{5\cos(\bar k\bar r)}{6\bar k}{\cal F}_2^{II}(t)-\frac{(2\cos(\bar k\bar r)^2-1)}{8\bar k^2\cos(\bar k\bar r)^2}\left[9\frac{\ddot a_0}{a_0}+9\left(\frac{\dot a_0}{a_0}\right)^2-32\pi G_6\delta\Lambda_6\right]\nonumber\\
&+&\frac{{\cal F}_4^{II}(t)}{\cos(\bar k\bar r)^2}
\eeqa
\beqa
Z^{III} &=& \frac{{\cal F}^{III}_1(t)}{\sin(\bar k\bar r)}-2\frac{\cos(\bar k\bar r)}{\bar k\sin(\bar k\bar r)}\left[\frac{\ddot a_0}{a_0}-\left(\frac{\dot a_0}{a_0}\right)^2\right]\\
W^{III} &=& {\cal F}^{III}_2(t)\sin(\bar k\bar r)\\
Y^{III} &=& -\frac{ \bar k\sin(\bar k\bar r)\cos(\bar k\bar r)}{8\pi G_6\beta\bar k_1}{\cal F}_2^{III}(t)+\frac{ \sin(\bar k\bar r)}{8\pi G_6\beta\bar k_1 }\left[3\frac{\ddot a_0}{a_0}+3\left(\frac{\dot a_0}{a_0}\right)^2-8\pi G_6\delta\Lambda_6\right]\\
X^{III} &=& \frac{5\cos(\bar k\bar r)}{6\bar k}{\cal F}_2^{III}(t)-\frac{(2\cos(\bar k\bar r)^2-1)}{8\bar k^2\cos(\bar k\bar r)^2}\left[9\frac{\ddot a_0}{a_0}+9\left(\frac{\dot a_0}{a_0}\right)^2-32\pi G_6\delta\Lambda_6\right]\nonumber\\
&+&\frac{{\cal F}_4^{III}(t)}{\cos(\bar k\bar r)^2}
\eeqa
\beqa
Z^{IV} &=& \frac{{\cal F}^{IV}_1(t)}{\sin(k\hat r)}-2\frac{\cos(k\hat r)}{k\sin(k\hat r)}\left[4\pi G_6(\rho_*+ p_*)+\frac{\ddot a_0}{a_0}-\left(\frac{\dot a_0}{a_0}\right)^2\right]\\
W^{IV} &=& {\cal F}^{IV}_2(t)\sin(k\hat r)-\frac{8\pi G_6\sin(k\hat r)\cos(k\hat r)}{k^3}{\cal P}_{*6}\\
\tilde p_{*5} &=& {\cal F}_3^{IV}(t)+\frac{\sigma\cos(k\hat r)}{k}{\cal F}_2^{IV}(t)-\frac{(k^2+8\pi G_6\sigma)(\cos(k\hat r)^2-\sin(k\hat r)^2)}{4k^4}{\cal P}_{*6}\\
Y^{IV} &=& \frac{(k^2+8\pi G_6\sigma+2(k^2-8\pi G_6\sigma)\cos(k\hat r)^2)\sin(k\hat r)}{4k^5\beta}{\cal P}_{*6}\nonumber\\
&-&\frac{(k^2-8\pi G_6\sigma)\sin(k\hat r)\cos(k\hat r)}{8\pi G_6\beta k^2}{\cal F}_2^{IV}(t)\nonumber\\
&+&\frac{\sin(k\hat r)}{8\pi G_6\beta k}\left[3\frac{\ddot a_0}{a_0}+3\left(\frac{\dot a_0}{a_0}\right)^2+8\pi G_6 {\cal F}_3^{IV}(t)-8\pi G_6\delta\Lambda_6\right]\\
X^{IV} &=& \frac{{\cal F}_4^{IV}(t)}{\cos(k\hat r)^2}+\frac{(5k^2-16\pi G_6\sigma)\cos(k\hat r)}{6k^3}{\cal F}_2^{IV}(t)\nonumber\\
&+&\frac{\pi G_6}{16k^6\cos(k\hat r)^2} \left[17k^2+16\pi G_6\sigma+32\cos(k\hat r)^2(k^2-4\pi G_6\sigma)\right.\nonumber\\
&-&\left.8\cos(k\hat r)^4(11k^2-16\pi G_6\sigma)\right]{\cal P}_{*6}\nonumber\\
&-&\frac{(2\cos(k\hat r)^2-1)}{8k^2\cos(k\hat r)^2} \left[9\frac{\ddot a_0}{a_0}+9\left(\frac{\dot a_0}{a_0}\right)^2-4\pi G_6(\rho_*-3 p_*+8\delta\Lambda_6-4{\cal F}_3^{IV}(t))\right]\\
 \tilde {p_*}^r_t &=& -3\frac{\cos(k\hat r)}{k\sin(k\hat r)}\frac{\dot a_0}{a_0}(\rho_*+ p_*)+\frac{d}{dt}\left[-\frac{9\sigma(1+2\cos(k\hat r)^2)}{8k^3\sin(k\hat r)\cos(k\hat r)}\left(\frac{\ddot a_0}{a_0}+\left(\frac{\dot a_0}{a_0}\right)^2\right)\right.\nonumber\\
&+&\left.\frac{\sigma\pi G_6-2\cos(k\hat r)^2(k^2-\sigma\pi G_6)}{2k^3\sin(k\hat r)\cos(k\hat r)}\rho_*-\frac{\sigma\pi G_6(1+2\cos(k\hat r)^2)}{k^3\sin(k\hat r)\cos(k\hat r)}\left(\frac{3}{2} p_*+2{\cal F}_3^{IV}(t)\right)\right.\nonumber\\
&+&\left.\frac{\sigma\pi G_6} {48k^7\sin(k\hat r)\cos(k\hat r)} \left\{-3(17k^2+16\sigma\pi G_6)+96\cos(k\hat r)^2(k^2-4\sigma\pi G_6)\right.\right.\nonumber\\
&-&\left.\left.8\cos(k\hat r)^4(11k^2-16\sigma\pi G_6)\right\}{\cal P}_{*6}\right.\nonumber\\
&+&\left.\frac{(5k^2-16\pi G_6\sigma)\sigma\cos(k\hat r)^2}{12k^4\sin(k\hat r)}{\cal F}_2^{IV}(t)-\frac{\sigma}{k\sin(k\hat r)\cos(k\hat r)}{\cal F}_4^{IV}(t)\right]+\frac{{\cal F}_5^{IV}(t)}{\sin(k\hat r)}
\eeqa
These solutions solve the equations of motion (\ref{TTminXX}-\ref{CONS1}).  However, we still need to impose boundary conditions.

Before doing so, we will specialize to a gauge where $E_1=B_1=0$.  We can then use the above results 
to find $C_1(r,t)$ and $A^{(1)}_{\theta}(r,t)$.  (The explicit solutions for $A_1(r,t)$ and $N_1(r,t)$ are of no 
particular interest, so we won't bother writing them explicitly).
\beqa
C_1^I &=& {\cal F}_4^I(t)+\frac{k\cos(kr)}{\sin(kr)}{\cal F}_6^I(t)+\frac{(5k^2-16\sigma\pi G_6)\cos(kr)}{6k^3}{\cal F}_2^I(t)\nonumber\\
&+&\frac{\pi G_6}{16k^6\sin(kr)}\left[\sin(kr)(17k^2+16\sigma\pi G_6-4\cos(kr)^2(11k^2-16\sigma\pi G_6))\right.\nonumber\\
&-&\left.4(3k^2+16\sigma\pi G_6)kr\cos(kr)\right]{\cal P}_6\nonumber\\
&-&\frac{2kr\cot(kr)-1}{8k^2}\left[9\frac{\ddot a_0}{a_0}+9\left(\frac{\dot a_0}{a_0}\right)^2-4\pi G_6(\rho-3 p+8\delta\Lambda_6-4{\cal F}_3^{I}(t))\right]\\
C_1^{II} &=& {\cal F}_4^{II}(t)+\frac{\bar k\cos(\bar k\bar r)}{\sin(\bar k\bar r)}{\cal F}_6^{II}(t)+\frac{5\cos(\bar k\bar r)}{6\bar k}{\cal F}_2^{II}(t)\nonumber\\
&-&\frac{2\bar kr\cot(\bar k\bar r)-1}{8\bar k^2}\left[9\frac{\ddot a_0}{a_0}+9\left(\frac{\dot a_0}{a_0}\right)^2-32\pi G_6\delta\Lambda_6\right]\\
C_1^{III} &=& {\cal F}_4^{III}(t)+\frac{\bar k\cos(\bar k\bar r)}{\sin(\bar k\bar r)}{\cal F}_6^{III}(t)+\frac{5\cos(\bar k\bar r)}{6\bar k}{\cal F}_2^{III}(t)\nonumber\\
&-&\frac{2\bar kr\cot(\bar k\bar r)-1}{8\bar k^2}\left[9\frac{\ddot a_0}{a_0}+9\left(\frac{\dot a_0}{a_0}\right)^2-32\pi G_6\delta\Lambda_6\right]\\
C_1^{IV} &=& {\cal F}_4^{IV}(t)+\frac{k\cos(k\hat r)}{\sin(k\hat r)}{\cal F}_6^{IV}(t)+\frac{(5k^2-16\sigma\pi G_6)\cos(k\hat r)}{6k^3}{\cal F}_2^{IV}(t)\nonumber\\
&+&\frac{\pi G_6}{16k^6\sin(k\hat r)}\left[\sin(k\hat r)(17k^2+16\sigma\pi G_6-4\cos(k\hat r)^2(11k^2-16\sigma\pi G_6))\right.\nonumber\\
&-&\left.4((3k^2+16\sigma\pi G_6)kr+(11k^2-16\sigma\pi G_6)k\phi)\cos(k\hat r)\right]{\cal P}_{*6}\nonumber\\
&-&\frac{2kr\cot(k\hat r)-1}{8k^2}\left[9\frac{\ddot a_0}{a_0}+9\left(\frac{\dot a_0}{a_0}\right)^2-4\pi G_6(\rho_*-3 p_*+8\delta\Lambda_6-4{\cal F}_3^{IV}(t))\right]
\eeqa
\beqa
{A^{(1)}_{\theta}}^I &=& {\cal F}_7^I(t)-\frac{\beta\sin(kr)}{k}{\cal F}_6^I(t)+\frac{\beta\cos(kr)}{k^2}{\cal F}_4^I(t)
\nonumber\\
&+&\frac{(3k^2(k^2-8\sigma\pi G_6)+4\beta^2\pi G_6(5k^2-16\sigma\pi G_6))\cos(kr)^2}{48\pi G_6\beta k^5}{\cal F}_2^I(t)\nonumber\\
&+&\frac{(4\beta^2\pi G_6 kr\sin(kr)-\cos(kr)(k^2-6\beta^2\pi G_6))}{k^4\beta}{\cal F}_3^I(t)\nonumber\\
&+&\frac{1}{48k^8\beta}\left[(-8k^2(k^2-8\sigma\pi G_6)-4\beta^2\pi G_6(11k^2-16\sigma\pi G_6))\cos(kr)^3\right.\nonumber\\
&-&\left.(12k^2(k^2+8\sigma\pi G_6)-3\beta^2\pi G_6(29k^2+80\sigma\pi G_6)))\cos(kr)\right.\nonumber\\
&+&\left.12\pi G_6\beta^2(3k^2+16\sigma\pi G_6)kr\sin(kr)\right]{\cal P}_6(t)\nonumber\\
&+&\frac{3(6\beta^2\pi G_6kr\sin(kr)-\cos(kr)(k^2-9\beta^2\pi G_6))}{8k^4\beta\pi G_6}\left[\frac{\ddot a_0}{a_0}+\left(\frac{\dot a_0}{a_0}\right)^2\right]\nonumber\\
&-&\frac{\beta\pi G_6(2kr\sin(kr)+3\cos(kr))}{2k^4}(\rho-3 p)\nonumber\\
&-&\frac{8\beta^2\pi G_6kr\sin(kr)-\cos(kr)(k^2-12\beta^2\pi G_6)}{k^4\beta}\delta\Lambda_6\\
{A^{(1)}_{\theta}}^{II} &=& {\cal F}_7^{II}(t)-\frac{\beta\sin(\bar k\bar r)}{\bar k_1}{\cal F}_6^{II}(t)+\frac{\beta\cos(\bar k\bar r)}{\bar k\bar k_1}{\cal F}_4^{II}(t)
\nonumber\\
&+&\frac{-3\bar k^2+(3\bar k^2+20\beta^2\pi G_6)\cos(\bar k\bar r)^2}{48\pi G_6\beta \bar k^2\bar k_1}{\cal F}_2^{II}(t)\nonumber\\
&+&\frac{3(6\beta^2\pi G_6\bar kr\sin(\bar k\bar r)-\cos(\bar k\bar r)(\bar k^2-9\beta^2\pi G_6))}{8\bar k^3\bar k_1\beta\pi G_6}\left[\frac{\ddot a_0}{a_0}+\left(\frac{\dot a_0}{a_0}\right)^2\right]\nonumber\\
&-&\frac{8\beta^2\pi G_6\bar kr\sin(\bar k\bar r)-\cos(\bar k\bar r)(\bar k^2-12\beta^2\pi G_6)}{\bar k^3\bar k_1\beta}\delta\Lambda_6\\
{A^{(1)}_{\theta}}^{III} &=& {\cal F}_7^{III}(t)-\frac{\beta\sin(\bar k\bar r)}{\bar k_1}{\cal F}_6^{III}(t)+\frac{\beta\cos(\bar k\bar r)}{\bar k\bar k_1}{\cal F}_4^{III}(t)
\nonumber\\
&+&\frac{-3\bar k^2+(3\bar k^2+20\beta^2\pi G_6)\cos(\bar k\bar r)^2}{48\pi G_6\beta \bar k^2\bar k_1}{\cal F}_2^{III}(t)\nonumber\\
&+&\frac{3(6\beta^2\pi G_6\bar kr\sin(\bar k\bar r)-\cos(\bar k\bar r)(\bar k^2-9\beta^2\pi G_6))}{8\bar k^3\bar k_1\beta\pi G_6}\left[\frac{\ddot a_0}{a_0}+\left(\frac{\dot a_0}{a_0}\right)^2\right]\nonumber\\
&-&\frac{8\beta^2\pi G_6\bar kr\sin(\bar k\bar r)-\cos(\bar k\bar r)(\bar k^2-12\beta^2\pi G_6)}{\bar k^3\bar k_1\beta}\delta\Lambda_6\\
{A^{(1)}_{\theta}}^{IV} &=& {\cal F}_7^{IV}(t)-\frac{\beta\sin(k\hat r)}{k}{\cal F}_6^{IV}(t)+\frac{\beta\cos(k\hat r)}{k^2}{\cal F}_4^{IV}(t)
\nonumber\\
&-&\frac{(3k^2(k^2-8\sigma\pi G_6)\tan(k\hat r)^2-4\beta^2\pi G_6(5k^2-16\sigma\pi G_6))\cos(k\hat r)^2}{48\pi G_6\beta k^5}{\cal F}_2^{IV}(t)\nonumber\\
&+&\frac{(4\beta^2\pi G_6 kr\sin(k\hat r)-\cos(k\hat r)(k^2-6\beta^2\pi G_6))}{k^4\beta}{\cal F}_3^{IV}(t)\nonumber\\
&+&\frac{1}{48k^8\beta}\left[(-8k^2(k^2-8\sigma\pi G_6)-4\beta^2\pi G_6(11k^2-16\sigma\pi G_6))\cos(k\hat r)^3\right.\nonumber\\
&-&\left.(12k^2(k^2+8\sigma\pi G_6)-3\beta^2\pi G_6(29k^2+80\sigma\pi G_6)))\cos(k\hat r)\right.\nonumber\\
&-&\left.12\pi G_6\beta^2k(r(-3k^2-16\sigma\pi G_6)+\phi(-11k^2+16\sigma\pi G_6))\sin(k\hat r)\right]{\cal P}_{*6}(t)\nonumber\\
&+&\frac{3(6\beta^2\pi G_6kr\sin(k\hat r)-\cos(k\hat r)(k^2-9\beta^2\pi G_6))}{8k^4\beta\pi G_6}\left[\frac{\ddot a_0}{a_0}+\left(\frac{\dot a_0}{a_0}\right)^2\right]\nonumber\\
&-&\frac{\beta\pi G_6(2kr\sin(k\hat r)+3\cos(k\hat r))}{2k^4}(\rho_*-3 p_*)\nonumber\\
&-&\frac{8\beta^2\pi G_6kr\sin(k\hat r)-\cos(k\hat r)(k^2-12\beta^2\pi G_6)}{k^4\beta}\delta\Lambda_6
\eeqa
The conditions we will impose on the solutions are the following:  they must all be regular at the poles, and 
must satisfy the jump conditions at the boundaries between the cores and the bulk.  Furthermore, all solutions 
must match smoothly across the equator separating the upper and lower core, with the exception of the gauge field 
perturbations, which as explained in the main text, must be related by a single valued gauge transformation.

The important point to watch out for comes from the equation of motion for the gauge invariant variable $X(r,t)$, 
eq.(\ref{OTHER}).  Indeed, this equation includes one dimensional delta functions coming from expanding the 
step function in terms of $\Delta r_0(t)$ and $\Delta r_*(t)$.  Concretely, this tells us that the radial 
derivative of $C_1$ is not smooth across the boundaries, but rather obeys
\beqa
\lim_{\epsilon\rightarrow 0}\left[{C_1^{II}}'(r_0+\epsilon,t)-{C_1^I}'(r_0-\epsilon,t)\right] &=& -8\pi G_6\sigma \Delta r_0(t)\\
\lim_{\epsilon\rightarrow 0}\left[{C_1^{IV}}'(r_*+\epsilon,t)-{C_1^{III}}'(r_*-\epsilon,t)\right] &=& 8\pi G_6\sigma \Delta r_*(t)
\eeqa
which we will regard as solving for $\Delta r_0(t)$ and $\Delta r_*(t)$ once all other functions have 
been set by the other boundary conditions.  It is interesting to note that the appearance of one dimensional 
delta functions in our solutions is in line with the arguments recently presented in \cite{Kanno}.  

Once all boundary conditions have been imposed, we have explicit expressions for all the integration 
``constants'' ${\cal F}_i(t)$ and we also have three additional equations which turn out to give 
one of the Friedmann equations (involving the combination $\ddot a_0/a_0-(\dot a_0/a_0)^2$) and two conservation of energy equations (one for the energy density on 
each brane).  Consistency between these three expressions allows us to recover the expression for the Hubble rate 
$(\dot a_0/a_0)^2$ in terms of the energy densities.  The expression written in terms of the six dimensional 
quantities are not very instructive however, so we will not write them out.  

Armed with full solutions, we can now work out the effective four dimensional quantities, which as explained 
in the main text, are given by
\beqa
\sigma^{(4)} &=& 2\pi \int_{0}^{r_0}e^{C_0(r)}\sigma dr = 2\pi \int_{\pi/k-\phi-r_0}^{ \pi/k-\phi } e^{C_0(r)}\sigma dr\\
\rho^{(4)}(t) &=& 2\pi\int_{0}^{r_0}e^{C_0(r)}\rho(t) dr + 2\pi\int_{0}^{r_0}e^{C_0(r)}\sigma C_1(r,t) dr +2\pi\sigma e^{C_0(r_0)}\Delta r_0(t)\\
p^{(4)}(t) &=& 2\pi\int_{0}^{r_0}e^{C_0(r)}p(t) dr - 2\pi\int_{0}^{r_0}e^{C_0(r)}\sigma C_1(r,t) dr -2\pi\sigma e^{C_0(r_0)}\Delta r_0(t)\\
\rho_*^{(4)}(t) &=& 2\pi\int_{\pi/k-\phi-r_0}^{ \pi/k-\phi } e^{C_0(r)}\rho_*(t) dr + 2\pi\int_{\pi/k-\phi-r_0}^{ \pi/k-\phi } e^{C_0(r)}\sigma C_1(r,t) dr \nonumber\\&-&2\pi\sigma e^{C_0(r_*)}\Delta r_*(t)\\
p_*^{(4)}(t) &=& 2\pi\int_{\pi/k-\phi-r_0}^{ \pi/k-\phi } e^{C_0(r)}p_*(t) dr - 2\pi\int_{\pi/k-\phi-r_0}^{ \pi/k-\phi } e^{C_0(r)}\sigma C_1(r,t) dr \nonumber\\&+&2\pi\sigma e^{C_0(r_*)}\Delta r_*(t).
\eeqa
Plugging these into the expressions for the Friedmann equations and conservation of energy allows us to 
derive eqs.(\ref{fried1}-\ref{cons2}).

%%%%%%%%%%%%%%%%%%%%%%%%%%%%%%%%%%%%%%%%%%%%%%%%%%%%%%%%%%%%%%%%%%%%%%%%%%%%%%%%%%%%%%%%%%%%%%%%%%%%
%---------------------------------------------------------------------------------------------------
%%%%%%%%%%%%%%%%%%%%%%%%%%%%%%%%%%%%%%%%%%%%%%%%%%%%%%%%%%%%%%%%%%%%%%%%%%%%%%%%%%%%%%%%%%%%%%%%%%%%

\end{document}